\documentclass{article}
\usepackage{main}
\usepackage{amsmath,amssymb}
\usepackage{graphicx}
\usepackage{natbib}
\usepackage{caption}
\usepackage{newfloat}
\usepackage{listings}
\usepackage{xcolor}

\usepackage{multirow}
\usepackage{multicol}
\usepackage{booktabs}
\usepackage{arydshln}
\usepackage{color}
\usepackage{makecell}
\usepackage{subcaption}
\usepackage{tabularx}
\usepackage{colortbl}
\usepackage{pifont}
\usepackage{wrapfig}

\definecolor{lightgreen}{RGB}{150, 250, 170}
\definecolor{lightred}{RGB}{255, 100, 100}
\definecolor{yuangreen}{HTML}{00CC70}
\hypersetup{allcolors=yuangreen}
\usepackage[most]{tcolorbox}
\usepackage{float}
\usepackage{xspace}
\usepackage{enumitem}
\tcbset{
  promptbox/.style={
    width=0.95\textwidth,
    colback=blue!5!gray!5,
    colframe=blue!30!gray,
    colbacktitle=blue!5!gray!5!,
    fonttitle=\bfseries\color{black},
    coltitle=black,
    boxrule=0.3mm,
    arc=3mm,
    top=6pt,
    bottom=6pt,
    left=10pt,
    right=10pt,
    enhanced,
    center,
    attach boxed title to top left={xshift=5mm, yshift=-3mm},
    boxed title style={
      arc=2mm,
      colframe=blue!30!gray,
      colback=blue!15!gray!15!
    },
    title filled,
    drop shadow={xshift=0.5mm, yshift=-0.5mm, fill=gray!30}
  }
}

\newtcolorbox{PromptBox}[2][]{promptbox, title=#2, #1}

\newcommand{\ours}{\textsc{LARF}}

\newcommand{\figref}[1]{Figure~\ref{#1}}

\title{\ours: Learning-aligned Reasoning Framework for Search Relevance}
\author{
Tian Tang\textsuperscript{1,{\color{yuangreen}*}} \quad Guo Chen\textsuperscript{1,2,{\color{yuangreen}*}} \quad Reed Li\textsuperscript{1,{\color{yuangreen}\dag}} \quad Hao Wang\textsuperscript{1} \quad Yu Lu\textsuperscript{1,{\color{yuangreen}\dag}}\\[6pt] Haijin Liang\textsuperscript{1} \quad
Haibo Shi\textsuperscript{1}\\[6pt]
\textsuperscript{1}Yuanbao Team, Tencent \quad \textsuperscript{2}Southwest University \\
}
\newcommand{\paperrunningtitle}{LARF: Learning-aligned Reasoning Framework for Search Relevance}
\newlength{\barwidth}
  \makeatletter
  \newcommand\blfootnote[1]{\begingroup
    \renewcommand\thefootnote{}\renewcommand{\@makefntext}[1]{\parindent=0pt\relax
      \noindent ##1}\footnote{#1}\addtocounter{footnote}{-1}\endgroup
  }
  \makeatother
\begin{document}

\maketitle
\blfootnote{\textsuperscript{{\color{yuangreen}*}}Equal Contribution. \textsuperscript{{\color{yuangreen}\dag}}Corresponding author: luyuresearch@foxmail.com, reedsli@tencent.com}

\begin{abstract}
Query-document relevance estimation is a core component of AI-native search systems, where a model must assess whether retrieved documents satisfy users' information needs under complex, multi-dimensional criteria. However, directly applying large language models (LLMs) to this task remains challenging because conventional supervised fine-tuning (SFT) provides limited supervision for domain-specific judgment rationales, while standard trajectory-level reinforcement learning (RL) objectives do not explicitly enforce margin separation among adjacent ordinal relevance labels. Moreover, although such reasoning can be useful during training, chain-of-thought (CoT) reasoning is impractical for latency-sensitive online ranking. These challenges make it difficult to resolve ambiguous decision boundaries. We propose a three-stage framework \ours~(Learning-Aligned Reasoning Framework for Search Relevance): it first performs cold-start SFT on strategically selected samples with multi-agent guided CoT rationales, then applies discriminative GRPO with attribute-wise, list-wise, and rule-consistency rewards together with a margin loss on the terminal relevance token, and finally distills the teacher into a dual-head student whose classification heads are retained for non-autoregressive inference. On a human-annotated web search relevance benchmark with approximately 300K training and 16K test instances, LARF-8B improves six-class classification accuracy by 5.78 percentage points and NDCG by 5.05 points, while the classification-only serving path reduces inference latency by 82.9\%. These results indicate that aligning structured reasoning, terminal decision optimization, and classification-head distillation can provide an effective accuracy--efficiency trade-off for industrial search relevance modeling. 
\end{abstract}
\thispagestyle{firstpage}

\section{Introduction}
Recent advances in large language models (LLMs) are reshaping information retrieval (IR) from conventional document matching toward a broader LLM-enhanced paradigm, encompassing generative retrieval, LLM-based reranking, retrieval-augmented generation, and search agents~\citep{zhu2025large,li2025matching}. Building on these capabilities, AI-native search systems integrate retrieval, reasoning, and answer generation to support complex and open-ended information needs; representative products include ChatGPT, Google Search with Gemini, Doubao and Yuanbao~\citep{openai2024chatgptsearch,google2025geminisearch,tencent2026yuanbao,bytedance2026doubao}. In such systems, search relevance remains a central determinant of search quality, as it governs whether retrieved documents provide useful evidence for satisfying users' underlying information needs, thereby affecting downstream answer quality, retrieval precision, and user satisfaction~\citep{mao2016does,tang2025lref}. AI-native search systems synthesize retrieved evidence into direct answers, making relevance estimation essential to both ranking quality and the faithfulness and usefulness of generated responses.

However, a central unresolved question is how to build LLM-based relevance models that can learn domain-specific judgment rules, make reliable fine-grained relevance decisions, and remain efficient enough for online search systems.
Supervised fine-tuning (SFT) maps query-document pairs to relevance labels~\citep{yuan2023multi,qiao2019understanding,wiseman2016sequence}. Label-only SFT learns input-output mappings but provides limited supervision for the rationales underlying relevance judgments~\citep{xia2024less,chang2024survey,lu2025rethinking}.
Conventional reinforcement learning (RL) methods such as GRPO primarily optimize trajectory-level rewards~\citep{zeng2026optimizing}, which creates two limitations for search relevance modeling. First, rewarding the entire reasoning trajectory can favor unnecessarily elaborate rationales, weakening the learning of concise discriminative representations~\citep{renze2024self}. Second, because the final relevance prediction is determined by a classification token (e.g., search relevance class), trajectory-level rewards provide no explicit fine-grained supervision for this decisive token. Models may still struggle to distinguish neighboring relevance grades~\citep{liu2025reasonrank}.
Although CoT combined with RL can improve relevance modeling, multi-step autoregressive generation incurs substantial latency~\citep{wei2022chain,chu2023survey,yeo2025demystifying,chen2026lir3ag}. This latency is often unacceptable for online systems.

Taken together, these limitations indicate that simply scaling label supervision or optimizing longer reasoning trajectories is insufficient for search relevance modeling. First, reasoning should serve as a training-time signal that exposes the structured criteria underlying relevance judgments, so that judgment rationales can be learned from a small strategically selected corpus. Second, the terminal ordinal decision should be optimized explicitly and along multiple relevance dimensions, rather than being left to a sparse trajectory-level reward. 
Third, autoregressive CoT generation increases serving latency and costs, motivating distillation into efficient students for real-time ranking~\citep{chen2023mcckdmulticotconsistentknowledge,eigner2024determinants,yang2024llm,xu2025dualheadreasoningdistillationimproving}. These needs call for a unified framework for multi-dimensional relevance learning, explicit decision optimization, and efficient serving.

In this paper, we propose \ours~(\textbf{L}earning-\textbf{A}ligned \textbf{R}easoning \textbf{F}ramework for Search Relevance), a three-stage framework that couples reasoning-oriented learning with efficient relevance serving: \textbf{(1) Cold-Start SFT with Guided CoT} constructs the training subset through stratified diversity sampling and hard-sample mining, allowing the model to learn from the most informative boundary cases, and then augments human relevance labels with multi-agent rationales that decompose judgments into diverse dimension; \textbf{(2) Discriminative GRPO with Margin Alignment} integrates point-wise, list-wise, and rule-consistency rewards to align generated rationales and relevance preferences, while applying a margin objective to explicitly separate the ground-truth relevance token from its strongest competing class~\citep{shao2024deepseekmathpushinglimitsmathematical,rosset2003margin}; \textbf{(3) Dual-head Knowledge Distillation} trains a shared backbone with multi-dimensional classification heads and an auxiliary generative reasoning head, and then performs efficient large-scale knowledge distillation using the classification heads~\citep{xu2025dualheadreasoningdistillationimproving}. Overall, \ours~separates structured reasoning learning, discriminative ordinal decision alignment, and efficient classification-only serving into a unified training-to-deployment framework for search relevance modeling.

We evaluate \ours~through extensive experiments. On a human-annotated web search relevance benchmark with approximately 300K training instances and 16K test instances, \ours-8B improves six-class accuracy from 45.20\% to 50.98\% and NDCG from 74.51 to 79.56. Moreover, the classification-only serving path reduces per-pair inference latency from 5.68 seconds to 0.97 seconds relative to autoregressive inference, which corresponds to a speedup of 5.86 times. In the Yuanbao AI Search A/B test, \ours~increases the high-satisfaction rate by 1.77\% and reduces the low-satisfaction rate by 3.22\%. These results demonstrate a practical accuracy--efficiency trade-off through structured reasoning during training and discriminative classification during serving.

The main contributions are summarized as follows:
\begin{itemize}
    \item We analyze LLM-based search relevance modeling through three joint challenges: structured relevance reasoning, cross-level decision alignment, and latency-sensitive online serving. We formalize search relevance as a multi-dimensional ordinal classification problem that integrates content quality, semantic relevance, source authority, and timeliness.

    \item We propose \ours~(\textbf{L}earning-\textbf{A}ligned \textbf{R}easoning \textbf{F}ramework for Search Relevance), a three-stage learning-aligned reasoning framework for search relevance modeling. \ours~combines: (i) Cold-Start SFT with Guided CoT, (ii) Discriminative GRPO with Margin Alignment for terminal decision optimization, and (iii) Dual-head Distillation for efficient non-autoregressive serving.

    \item We evaluate \ours~across extensive experiments, including offline evaluation, component-wise ablations, and online A/B testing, and demonstrate that \ours~consistently improves search relevance quality while reducing serving latency and yielding gains in user-satisfaction metrics.
\end{itemize}

\section{Related Work}

\subsection{LLMs for Search Relevance Modeling}

Search relevance modeling estimates the extent to which a candidate document satisfies a user's information need and is a core component of retrieval and ranking systems. Early neural approaches established learned semantic representations for query-document matching, moving beyond lexical matching toward discriminative relevance estimation~\citep{shen2014learning,koroteev2021bert}. With the emergence of large language models, subsequent studies have explored prompt-based relevance prediction, task-specific supervised adaptation, data construction, and reasoning-oriented training for search relevance modeling~\citep{zhang2025survey,xie2025logic}. Another line of work investigates LLMs as relevance assessors by comparing binary, graded, and preference-based judgments with human annotations, revealing both their potential for relevance assessment and their sensitivity to judgment formulations~\citep{faggioli2023perspectives,arabzadeh2025benchmarking}. In particular, using fine-grained relevance labels instead of binary decisions has been shown to improve the discrimination of partially relevant documents in LLM-based ranking~\citep{zhuang2024beyond}. Recent industrial studies further combine informative data selection, structured rationale construction, and preference-oriented optimization to improve relevance judgments in complex search scenarios~\citep{tang2025lref,lu2025lore}. These advances demonstrate that LLMs can model relevance beyond shallow semantic similarity; however, existing methods provide limited treatment of relevance as a multi-dimensional ordinal decision problem and do not fully unify structured relevance learning, direct optimization of the final relevance decision, and efficient online serving within a single framework.

\subsection{Reasoning-Enhanced LLM Ranking and Decision Alignment}

Reasoning-enhanced LLM ranking investigates how language models can use intermediate reasoning and alignment objectives to improve relevance judgments for complex queries and ambiguous candidate documents. Foundational studies show that prompting LLMs to generate intermediate reasoning steps can improve performance on tasks requiring multi-step inference~\citep{wei2022chain}. In parallel, LLM-based rerankers have been developed under pointwise, pairwise, listwise, and setwise decision formulations, demonstrating different effectiveness--efficiency trade-offs in comparing and ordering candidate documents~\citep{qin2024large,luo2024prp,zhuang2024setwise}. Subsequent research examines whether explicit reasoning can benefit document ranking and introduces reasoning-aware training schemes for improving ranking quality and interpretability~\citep{lu2025rethinking,zhuang2025rank,liu2025reasonrank}. Recent approaches further employ reinforcement learning to optimize reasoning trajectories, ranking preferences, and rule-consistent model behavior through outcome-based or preference-based reward signals~\citep{zeng2026optimizing,guo2026reward,liu2026gdpo,zhang2025ratt}. Another related direction transfers training-time rationale supervision into efficient prediction models, allowing reasoning to improve learned representations without requiring autoregressive rationale generation at inference time~\citep{chen2023mcckdmulticotconsistentknowledge,eigner2024determinants,yang2024llm,xu2025dualheadreasoningdistillationimproving}. Nevertheless, existing approaches generally optimize reasoning trajectories or relative ranking outcomes without directly modeling the fine-grained ordinal boundary of the terminal relevance decision, and they do not jointly address multi-dimensional relevance criteria, terminal decision alignment, and latency-sensitive classification-only serving.

\section{Preliminaries}
This section formalizes the multi-dimensional query-document search relevance matching problem and introduces the structured prompt syntax for \ours~training.

\subsection{Multi-Dimensional Problem Formulation}
\label{sec:problem_formulation}
Compared to relying solely on document relevance based on embedding vectors or word frequencies, integrating multi-dimensional information is crucial for determining document relevance. Therefore, we formulate search relevance modeling as a multi-dimensional ordinal classification problem. For a query $q$, each candidate document $d \in \mathcal{D}_q$ is associated with document content and auxiliary metadata $m_d$ (e.g., source and publication time). We denote the complete model input by $x_{q,d} = (q,d,m_d)$ and use $\mathcal{G}_q=\mathcal{D}_q$ to denote the candidate document group for $q$.

Each query-document pair is annotated with an ordered label vector $\mathbf{y}_{q,d} = \big(y_{q,d}^{(k)}\big)_{k\in\mathcal{K}}$, where $\mathcal{K}=\{\mathrm{sem},\mathrm{aut},\mathrm{qua},\mathrm{time},\mathrm{rel}\}$ indexes semantic relevance, source authority, content quality, timeliness, and overall relevance, respectively. Every dimension takes values in the common ordinal label space
\begin{equation}
    \mathcal{Y} = \{0,1,2,3,4,5\}, \qquad 0 \prec 1 \prec \cdots \prec 5,
\end{equation}
where the terminal label $y_{q,d}^{(\mathrm{rel})}\in\mathcal{Y}$ ranges from \textit{Completely Irrelevant} to \textit{Perfectly Relevant}. The four auxiliary dimensions provide structured evidence for the overall judgment; we do not assume a fixed deterministic aggregation rule because their relative importance depends on the query and document context. Content quality evaluates document completeness and structure; semantic relevance measures intent and topical alignment; source authority captures the credibility of the originating source; and timeliness evaluates whether the document satisfies the query's temporal intent.
The model predicts an ordinal label vector $\hat{\mathbf{y}}_{q,d}$ for each query-document pair $d\in\mathcal{D}_q$. Learning therefore involves three complementary objectives: accurately predicting each relevance dimension, preserving the relative ordering of documents under the same query, and explicitly separating the terminal relevance label from competing classes. These objectives motivate the attribute-wise reward, list-wise reward, and terminal-label margin loss.

\subsection{Data Structure and Prompt Construction}
We serialize each input $x_{q,d}$ into an instruction-based training instance, pairing an input prompt $P(x_{q,d})$ with a structured target $T_{q,d}=(R_{q,d},\mathbf{y}_{q,d})$. The prompt includes a \texttt{[System]} directive that assigns the role of a search relevance evaluator, the query-document content and auxiliary metadata, and a \texttt{[Constraint]} that requires assessment of content quality, semantic relevance, source authority, and timeliness, then gives the final search relevance class. The target first generates a dimension-wise rationale $R_{q,d}$ and then outputs the ordered relevance labels, ending with the terminal six-level label $y_{q,d}^{(\mathrm{rel})}$. This format makes the position of the terminal decision token explicit while retaining the structured evidence used to derive it. Our formulation is prompt-driven, and the complete prompt and target templates are provided in Appendix~\ref{app:prompt}.

\begin{figure*}
    \centering
    \includegraphics[width=1\linewidth, trim=0 1.5cm 0 0, clip]{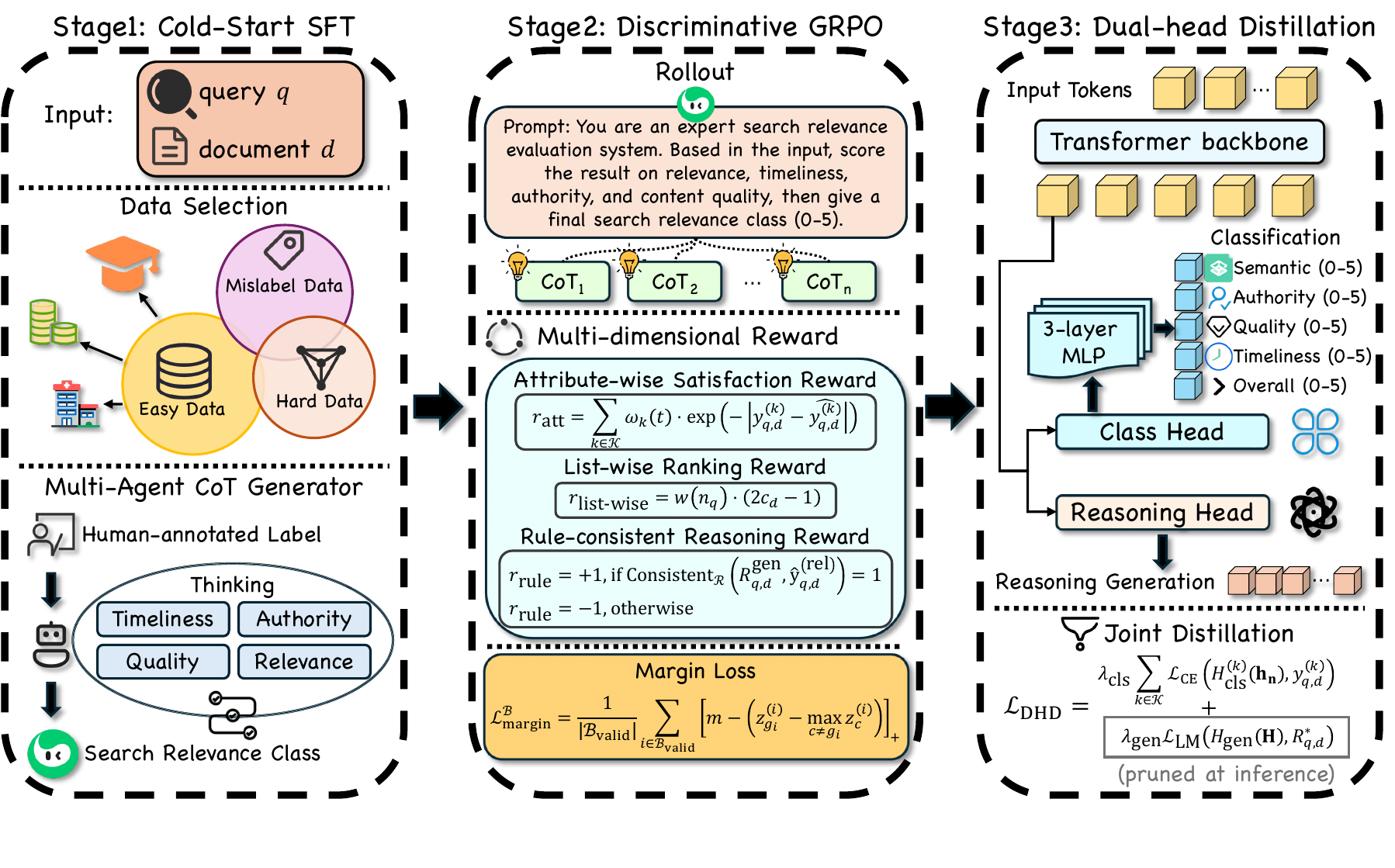}
    \caption{Overall workflow of \ours, including three stages: Cold-Start SFT with guided Chain-of-Thought, Discriminative GRPO with Margin Alignment, and Dual-head Knowledge Distillation.}
    \label{fig:fig1}
\end{figure*}
\section{Methodology}
\label{sec:methodology}
\subsection{\ours~Overview}
\label{sec:method_overview}
As shown in \figref{fig:fig1}, \ours~follows a three-stage training-to-serving pipeline. The first stage, \textbf{Cold-Start SFT with Guided CoT}, selects informative query-document pairs and augments their multi-dimensional relevance labels with structured rationales, providing reasoning-oriented supervision for the five relevance dimensions. The second stage, \textbf{Discriminative GRPO with Margin Alignment}, combines attribute-wise, list-wise, and rule-consistency rewards with a terminal-token margin loss, thereby coupling trajectory-level alignment with direct ordinal decision optimization. The third stage, \textbf{Dual-head Knowledge Distillation}, uses rationale generation as an auxiliary training signal for a shared backbone and retains only the multi-dimensional classification heads at inference time. The three stages respectively support structured relevance learning, terminal decision alignment, and non-autoregressive online serving.

\subsection{Cold-Start SFT with Guided Chain-of-Thought}
\label{sec:cold_start}
To establish a reasoning-oriented initialization for subsequent reinforcement learning, we introduce a Cold-Start Supervised Fine-Tuning (SFT) stage. Rather than indiscriminately scaling the training corpus, this stage uses a curated subset of informative instances paired with multi-agent synthesized Chain-of-Thought (CoT) rationales. Notably, empirical evaluations demonstrate that this cold-start paradigm, using only \textbf{one-third} of the data, achieves performance comparable to full-dataset tuning.

\subsubsection{Multi-Criteria Strategic Data Selection}
Our selection procedure combines coverage-oriented sampling with boundary-focused sampling. First, \textbf{Stratified Topic Coverage} samples across vertical domains (e.g., finance, entertainment, and healthcare) to broaden the input distribution. Second, \textbf{Relevance Category Density} retains, for each sampled query, candidate documents spanning as many distinct terminal relevance grades as available. We then blend these samples with targeted hard cases from two feedback channels: \textit{empirical error cases}, for which a baseline model disagrees with the annotated terminal label, and \textit{high-entropy instances}, for which the baseline assigns high predictive entropy $H\!\left(\hat{y}_{q,d}^{(\mathrm{rel})}\mid x_{q,d}\right)$, signaling intense model ambiguity during inference. The resulting subset combines ordinary examples with instances near uncertain ordinal boundaries.

\subsubsection{Multi-Agent CoT Rationale Synthesis}

To construct structured rationale targets for the cold-start corpus, we use an automated multi-agent synthesis pipeline that augments human relevance labels with dimension-wise analyses. The pipeline decomposes each relevance judgment through the following stages:

\begin{enumerate}
    \item \textbf{Dimensional Decomposition:} We deploy a suite of specialized LLM agents, each tasked with evaluating a specific sub-dimension of the search relevance task (e.g., Content Quality, Semantic Relevance, Source Authority, and Timeliness).
    \item \textbf{Independent Reasoning Generation:} Each agent independently assesses $x_{q,d}$, including the relevant auxiliary metadata, and produces a dimension-specific analysis with an intermediate ordinal assessment.
    \item \textbf{Consensus Harmonization:} A centralized coordinator aggregates the dimension-specific analyses into a unified rationale and a provisional overall assessment.
    \item \textbf{Label-Conditioned Target Construction:} The coordinator is conditioned on the human-annotated label vector $\mathbf{y}_{q,d}$ when constructing the SFT target, yielding $T_{q,d}=(R_{q,d},\mathbf{y}_{q,d})$. This conditioning is used only to construct label-consistent training targets; it does not constrain the rationales generated by the policy during the subsequent GRPO stage.
    \item \textbf{Hallucination Checking and Regeneration:} A prompt-based checker screens each synthesized rationale $R_{q,d}$ against its corresponding golden labels. Rationales that fail the check are regenerated, and those failing repeatedly are discarded. Only verified targets enter the cold-start corpus.
\end{enumerate}

The checker operates offline during corpus construction, screening targets that are already conditioned on gold labels. Ultimately, by coupling strategic data selection with this multi-agent CoT synthesis, our pipeline endows the model with a nuanced, multi-perspective understanding of relevance criteria. This robust initialization facilitates stable and efficient convergence during the subsequent reinforcement learning phase. 

\subsection{Discriminative GRPO with Margin Alignment}
While cold-start SFT establishes a reasoning foundation, it is limited by data scale and prone to memorizing reasoning paths. We therefore introduce a reinforcement-learning stage that evaluates sampled responses using multiple relevance signals while preserving direct supervision of the terminal decision. Furthermore, since search relevance is inherently a classification task, the model must also possess strong discriminative capabilities.

Discriminative GRPO with margin alignment combines a multi-dimensional reward system with a loss applied to the terminal relevance token. The reward components encourage attribute agreement, within-query ordering, and rule consistency, whereas the margin term explicitly separates the gold terminal label from its strongest competing label.

\subsubsection{Multi-dimensional Reward Decomposition}
\label{subsec:reward_decomp}
For each sampled response to $x_{q,d}$, we decompose the reward $r_{q,d}$ into three components: an attribute-wise reward, a list-wise ranking reward, and a rule-consistency reward.

\textbf{(1) Attribute-wise Relevance Reward ($r_{\mathrm{att}}$):} A holistic relevance label alone provides limited feedback about which relevance criterion is violated. We therefore use the five-dimensional label set $\mathcal{K}$ defined in Section~\ref{sec:problem_formulation} and compute an attribute-wise reward as follows:
\begin{equation}
r_{\mathrm{att}} = \sum_{k \in \mathcal{K}} \omega_k(t) \cdot \exp\!\left(-\left|y_{q,d}^{(k)} - \hat{y}_{q,d}^{(k)}\right|\right).
\end{equation}
Here, $y_{q,d}^{(k)}$ and $\hat{y}_{q,d}^{(k)}$ are the ground-truth and predicted ordinal labels for dimension $k$, respectively, and $\omega_k(t)$ is a dynamic dimension-specific weight. The exponential term transforms ordinal error into a smooth bounded reward. By jointly optimizing these auxiliary dimensions, the model develops a rigorous, multi-perspective understanding that significantly enhances the accuracy and robustness of the primary relevance prediction. 

\textbf{(2) List-wise Ranking Reward:} 
Standard LLM optimization paradigms typically treat search relevance as an isolated point-wise classification task. However, in real-world search scenarios, the relative ranking order—specifically, prioritizing highly relevant documents over less relevant ones—is equally critical. To explicitly optimize the relative preference order of multiple documents under the same query $q$, we design a query-grouped, list-wise reward. For a query group $\mathcal{G}_q$, let $P_d$, $N_d$, and $T_d$ denote the concordant, discordant, and tied pair counts for document $d$ relative to all other candidate documents $d' \in \mathcal{G}_q \setminus \{d\}$. These counts are derived by comparing the predicted terminal labels $\hat{y}^{(\mathrm{rel})}$ against the ground-truth terminal labels $y^{(\mathrm{rel})}$; pairs with equal ground-truth labels are excluded. We use a pairwise normalized relevance (PNR) contribution, defined as the normalized per-document concordance score:
\begin{equation}
    c_d = \frac{P_d + 0.5\,T_d}{P_d + N_d + T_d} \in [0,1].
\end{equation} 

To explicitly penalize ranking inversions and reinforce correct pairwise preferences, we apply a zero-centering transformation to the raw concordance rate $c_d$. By mapping $c_d$ around the random-guess threshold via $(2c_d - 1) \in [-1, 1]$, we establish a strict baseline: positive reinforcement ($>0$) is activated exclusively when the policy outperforms random permutations, while discordant rankings incur proportional penalties (down to $-1$). To further stabilize training, we introduce a group-size-aware confidence weight $w(n_q)$ that calibrates the reliability of the list-wise statistics based on the query group size $n_q = |\mathcal{G}_q|$, where $n_q$ denotes the number of candidate documents in the query group and $\bar{n}$ denotes the mean query group size in the training set. The weight $w(n_q)$ increases with the group size to reflect the reliability of pairwise ranking statistics, while the square-root scaling prevents large groups from dominating the reward. Ultimately, the final list-wise reward optimized via the GRPO algorithm is formulated as:
\begin{equation}
w(n_q)=\min\left(1,\sqrt{\frac{n_q-1}{\bar{n}-1}}\right),
\end{equation}

\begin{equation}
    r_{\mathrm{list-wise}} = w(n_q) \cdot (2c_d - 1).
\end{equation} 
\textbf{(3) Rule-consistent Reasoning Reward ($r_{\mathrm{rule}}$):} 
The label-conditioned SFT targets are not sufficient to ensure that rationales generated during RL remain consistent with annotation principles. We therefore define an explicit rule set $\mathcal{R}$, where each rule maps a verifiable relevance condition to compatible dimension-level or terminal labels (e.g., outdated content should not receive a high timeliness label). For each sampled rollout, a binary rule checker evaluates the compatibility of the generated rationale $R_{q,d}^{\mathrm{gen}}$ and terminal prediction $\hat{y}_{q,d}^{(\mathrm{rel})}$ with $\mathcal{R}$:
\begin{equation}
r_{\mathrm{rule}} = \begin{cases} 
+1, & \text{if } \mathrm{Consistent}_{\mathcal{R}}\big(R_{q,d}^{\mathrm{gen}}, \hat{y}_{q,d}^{(\mathrm{rel})}\big)=1, \\ 
-1, & \text{otherwise}.
\end{cases}
\end{equation}
By penalizing reasoning paths that violate human annotation principles, $r_{\mathrm{rule}}$ enhances the model's interpretability and robustness, particularly in boundary cases. The final reward for a sampled response is
\begin{equation}
r_{q,d}=r_{\mathrm{att}}+r_{\mathrm{list-wise}}+r_{\mathrm{rule}}. 
\end{equation}

\subsubsection{Margin-based Auxiliary Loss for Search Relevance Classification}
\label{subsec:margin_loss}
\paragraph{Motivation} 
In classification-centric search relevance ranking, the final prediction is strictly dictated by a single decision-digit token (e.g., the `3` in ``\texttt{search relevance: 3}''). Despite its paramount importance, conventional optimization paradigms fail to provide direct, token-level supervision for this crucial digit. Specifically, the standard GRPO policy gradient operates exclusively on trajectory-level rewards, yielding only a sparse and indirect signal. Similarly, CoT-augmented SFT losses lack the fine-grained optimization required to explicitly maximize the target category's probability and distance it from easily confusable classes. To address this vulnerability, we introduce a hinge-style margin loss applied specifically to the logits of this decision token. This auxiliary loss explicitly expands the decision boundary between the ground-truth digit and its strongest competitor, while its directional clipping property prevents overfitting on already well-separated samples. 

\paragraph{Formulation} 
Let $\mathbf{z}=(z_0,\dots,z_5)\in\mathbb{R}^{6}$ denote the logits restricted to the six candidate digit tokens at the designated terminal-label position, and let $g=y_{q,d}^{(\mathrm{rel})}\in\mathcal{Y}$ be the ground-truth terminal relevance label. We define the empirical classification margin $\Delta(\mathbf{z}, g)$ as:
\begin{equation}
\Delta(\mathbf{z}, g) \;=\; z_{g} \;-\; \max_{c \neq g}\, z_{c},
\end{equation}
which remains strictly positive if and only if the token is correctly predicted. The per-sample auxiliary objective adopts a hinge formulation with a target margin hyperparameter $m>0$ (default $m=1.0$):
\begin{equation}
\mathcal{L}_{\text{margin}}(\mathbf{z}, g) \;=\; \big[\, m - \Delta(\mathbf{z}, g) \,\big]_{+} \;=\; \max\!\Big(0,\; m - \big(z_{g} - \max_{c \neq g} z_{c}\big)\Big).
\end{equation}
For a micro-batch $\mathcal{B}$, we define $\mathcal{B}_{\mathrm{valid}}\subseteq\mathcal{B}$ as the subset of instances whose formatted targets contain exactly one identifiable terminal relevance-label position and whose gold label belongs to $\mathcal{Y}$. The aggregated auxiliary loss is computed as the empirical mean over this valid subset:
\begin{equation}
\mathcal{L}_{\text{margin}}^{\mathcal{B}} \;=\; \frac{1}{|\mathcal{B}_{\text{valid}}|} \sum_{i\in\mathcal{B}_{\text{valid}}} \big[\, m - \big(z^{(i)}_{g_i} - \max_{c\neq g_i} z^{(i)}_{c}\big) \,\big]_{+}.
\end{equation}

\paragraph{Integration with GRPO} 
This margin loss is integrated jointly into the main GRPO policy optimization objective with coefficient $\lambda_{\mathrm{margin}}$:
\begin{equation}
\mathcal{L}_{\mathrm{GRPO+margin}} = \mathcal{L}_{\mathrm{GRPO}} - \beta_{\mathrm{ent}} \mathcal{H}(\pi_{\theta}) + \lambda_{\mathrm{margin}} \mathcal{L}_{\mathrm{margin}}^{\mathcal{B}}.
\end{equation}
Here, $\mathcal{L}_{\mathrm{GRPO}}$ is the Group Relative Policy Optimization loss, in which the KL-divergence penalty relative to the reference policy $\pi_{\mathrm{ref}}$ is implicitly regularized within the reward function. $\mathcal{H}(\pi_{\theta})$ denotes policy entropy, scaled by $\beta_{\mathrm{ent}}$ to encourage exploration, and $\lambda_{\mathrm{margin}}$ balances the terminal-token margin loss. Gradients from this auxiliary term flow exclusively through the six digit-token logits at the terminal decision position.
\paragraph{Effect on Terminal Decisions.} 
The margin objective is designed to improve terminal classification through two mechanisms: 
(i)~\textbf{Adaptive Boundary Emphasis:} The $\max_{c\neq g} z_c$ operator identifies the strongest competing class for each sample. By forcing the optimization process to directly widen the most vulnerable decision boundary, it effectively resolves ambiguity between easily confused classes and significantly boosts accuracy on hard, borderline cases.
(ii)~\textbf{Direct Terminal-Token Supervision:} By anchoring this margin loss exclusively to the final search relevance category token, we ensure that the optimization of this critical decision point remains the highest priority throughout the RL phase. This provides a direct and highly efficient supervisory signal that maximizes the correct label's probability, overcoming the sparsity of trajectory-level rewards.

\subsection{Dual-head Knowledge Distillation}
\label{subsec:dual_head_distillation}

To reduce autoregressive CoT latency in industrial deployment, we further train a dual-head teacher initialized from the GRPO-trained model. It uses rationale supervision during training and classification-only inference at deployment, aiming to retain RL-acquired reasoning capabilities without generating multi-step rationales~\citep{xu2025dualheadreasoningdistillationimproving}.

The model comprises a unified, shared transformer backbone equipped with two task-specific parallel prediction layers: a discriminative classification module $H_{\mathrm{cls}}$ and a generative reasoning head $H_{\mathrm{gen}}$. Rather than predicting only overall relevance, we instantiate the classification module as five sub-heads $\{H_{\mathrm{cls}}^{(k)}\}_{k\in\mathcal{K}}$. Each sub-head predicts one component of $\mathbf{y}_{q,d}$. The joint training objective is
\begin{equation}
\mathcal{L}_{\mathrm{DHD}} = \lambda_{\mathrm{cls}} \sum_{k\in\mathcal{K}} \mathcal{L}_{\mathrm{CE}}\big(H_{\mathrm{cls}}^{(k)}(\mathbf{h}_n), y_{q,d}^{(k)}\big) + \lambda_{\mathrm{gen}} \mathcal{L}_{\mathrm{LM}}\big(H_{\mathrm{gen}}(\mathbf{H}), R_{q,d}^{*}\big),
\end{equation}
where $\mathbf{h}_n \in \mathbb{R}^d$ is the hidden state of the final input token, $\mathbf{H}$ is the sequence hidden-state matrix, and $R_{q,d}^{*}$ is the rationale generated by the GRPO-trained teacher for $x_{q,d}$. The coefficients $\lambda_{\mathrm{cls}}$ and $\lambda_{\mathrm{gen}}$ balance classification and rationale supervision, respectively. The generative reasoning head is an auxiliary training-time objective intended to encourage the shared backbone to encode reasoning-relevant information.

When distilling the search relevance model into the student model for online ranking, the generative reasoning head $H_{\mathrm{gen}}$ is entirely discarded. The optimized teacher model executes a single, non-autoregressive forward pass exclusively through the multi-dimensional classification heads $H_{\mathrm{cls}}$. This architectural removal delivers the high-accuracy semantic alignment of a reasoning-capable large language model while completely bypassing the latency overhead associated with token-by-token autoregressive decoding. 

\section{Experiments}

\begin{table}[t]
  \centering
    \resizebox{\linewidth}{!}{
      \begin{tabular}{l|cccc|ccc}
                \toprule
                \multirow{2}{*}{\textbf{Models}} &
                \multicolumn{4}{c|}{\textbf{Fine-grained Relevance Hierarchy}} &
                \multicolumn{3}{c}{\textbf{Coarse-grained Alignment}} \\
                & \textbf{NDCG} & \textbf{Macro F1} & \textbf{Weighted F1} & \textbf{Acc.}
                & \textbf{Prec.} & \textbf{Rec.} & \textbf{F1} \\
                \midrule
                Qwen3-Reranker-4B                                   & 72.23 & 34.10 & 40.80 & 42.30 & 61.23 & 66.93 & 63.95 \\
                Qwen3-Reranker-8B                                   & 74.20 & 36.50 & 42.30 & 44.10 & 64.50 & 68.10 & 66.25 \\
                \midrule
                Qwen3-1.7B-Base                                    & 70.15 & 31.20 & 38.50 & 40.10 & 58.40 & 63.50 & 60.85 \\
                Qwen3-8B-Instruct                                      & 74.51 & 37.40 & 43.15 & 45.20 & 68.30 & 77.20 & 72.48 \\
                Qwen3-235B-A22B-Instruct                           & 75.87 & 38.10 & 44.20 & 45.61 & 70.50 & 79.10 & 74.54 \\
                \midrule
                \ours-1.7B \textit{deployed student}               & 75.20 & 38.05 & 43.90 & 45.80 & 69.10 & 78.30 & 73.41 \\
                \midrule
                \ours-8B \textit{cold-start SFT}                   & 76.10 & 40.25 & 45.30 & 46.12 & 71.40 & 81.15 & 75.95 \\
                \ours-8B \textit{cold-start SFT + GRPO}            & 78.40 & 42.15 & 48.20 & 49.35 & 73.15 & 84.30 & 78.33 \\
                \ours-8B \textit{cold-start SFT + GRPO + DH}       & 79.56 & 43.68 & 50.13 & 50.98 & 74.82 & \textbf{86.45} & 80.22 \\
                \ours-235B \textit{cold-start SFT + GRPO + DH}      & \textbf{80.03} & \textbf{45.76} & \textbf{52.39} & \textbf{53.23} & \textbf{79.99} & 85.92 & \textbf{82.85} \\
                \bottomrule
        \end{tabular}
    }
\caption{Overall offline performance across two semantic granularities. \textbf{DH} denotes dual-head distillation. The best value in each column is shown in bold.}
\label{tab:experiment_overall}
\end{table}

\subsection{Datasets and Baselines}
\label{sec:Dataset_Baselines}
To evaluate our framework, we curate a high-quality, human-annotated web search relevance benchmark comprising a \textbf{Training Set} ($\sim$300K samples) and a \textbf{Test Set} ($\sim$16K samples). The training data features a hierarchical annotation schema—combining a primary relevance score with four sub-dimensions (semantic relevance, timeliness, content quality, and source authority)—to enable robust multi-dimensional supervision. Meanwhile, the test set is deliberately curated with hard negatives and challenging edge cases (e.g., missing tokens, spelling errors, and complex semantic variants) to rigorously assess generalization boundaries. 

We compare our proposed framework with two primary baseline paradigms: (1) Non-Reasoning Rerankers Qwen3-Reranker-4B and Qwen3-Reranker-8B~\citep{zhang2025qwen3}, which directly predict relevance scores via a single forward pass without intermediate rationales, and (2) Reasoning-Based LLMs Qwen3-8B-Instruct and Qwen3-235B-A22B-Instruct~\citep{yang2025qwen3}, which autoregressively generate rationales before predicting the final label. Models fine-tuned under our framework using these generative backbones are denoted as \textbf{LARF}. All LARF variants share the backbone, training corpus, and decoding configuration of their corresponding baselines, so that reported differences are attributable to the framework rather than to model capacity.

\subsection{Evaluation Metrics}
For offline evaluation, we report the standard Normalized Discounted Cumulative Gain (NDCG) to measure list-wise ranking quality \citep{wang2013theoretical}. To assess classification performance, we map search relevance into a six-tier grading system: \textit{Completely Irrelevant} (with safety violations), \textit{Irrelevant}, \textit{Partially Relevant}, \textit{Basically Relevant}, \textit{Highly Relevant}, and \textit{Perfectly Relevant}. Based on this schema, we introduce two diagnostic metrics: First, \textbf{Fine-grained Relevance Hierarchy} measures exact match accuracy, Macro F1, and Weighted F1 across all six levels, reflecting the model's sensitivity to subtle search relevance fluctuations. Second, \textbf{Coarse-grained Relevance Alignment} collapses the six levels into binary buckets (\textit{Irrelevant} and \textit{Relevant}) to evaluate macro-level alignment, boundary tolerance, and overall ranking stability. 

\subsection{Overall Performance}
Table~\ref{tab:experiment_overall} shows that \ours~primarily improves the ability to distinguish adjacent relevance grades rather than merely ranking clearly relevant documents higher. \ours-8B raises fine-grained accuracy from 45.20\% to 50.98\% and NDCG from 74.51 to 79.56 over its Qwen3-8B backbone, and exceeds the strongest non-reasoning reranker, Qwen3-Reranker-8B, by 6.88 points of accuracy and 13.97 points of coarse F1. More strikingly, \ours-8B outperforms the nearly 30-times-larger Qwen3-235B-A22B-Instruct base model across all metrics, demonstrating that these gains stem from the proposed learning framework rather than increased model capacity. For the best overall performance, we use the 235B model as the backbone and implement the full \ours~framework, achieving the highest results in the table: an accuracy of 53.23\% and an NDCG of 80.03.
The three stages contribute in the manner their designs predict. Cold-start SFT with guided chain-of-thought reaches 46.12\% accuracy and 76.10 NDCG. This result is primarily attributed to supervision of the reasoning process. Discriminative GRPO with margin alignment then produces the largest increment of \ours, 3.23 points of accuracy against 0.92 for the preceding stage, confirming our assumption that the dominant difficulty lies in aligning the terminal decision with ambiguous human boundaries rather than in acquiring surface-level semantic matching. Dual-head distillation adds a further 1.63 points to 50.98\% while removing autoregressive decoding from the online path, so efficiency is obtained without conceding accuracy. Coarse-grained results preserve the same stage ordering, with Recall rising from 77.20 to 86.45 and F1 from 72.48 to 80.22.

We note that the 8B variant retains the highest coarse Recall at 86.45 whereas the 235B variant attains the highest coarse Precision at 79.99, so the two trade error types at the binary boundary. In the opposite direction, the distilled \ours-1.7B student attains 45.80\% accuracy and 75.20 NDCG, retaining 89.8\% of its teacher's accuracy at roughly one-fifth of the parameters and improving over its own 1.7B backbone by 5.70 points. Most notably, this non-autoregressive student edges out the 8B base model by 0.60 points and the Qwen3-Reranker-8B baseline by 1.70 points, which is what makes deployment practical: the capability acquired through guided chain-of-thought and discriminative GRPO survives compression into a model small enough to serve online.
\begin{table}[t]
  \centering
    \resizebox{\linewidth}{!}{
      \begin{tabular}{l|ccc|ccc}
                \toprule
                \multirow{2}{*}{\textbf{Models}} &
                \multicolumn{3}{c|}{\textbf{Fine-grained Relevance Hierarchy}} &
                \multicolumn{3}{c}{\textbf{Coarse-grained Relevance Alignment}} \\
                & \textbf{Macro F1} & \textbf{Weighted F1} & \textbf{Accuracy}
                & \textbf{Precision} & \textbf{Recall} & \textbf{F1} \\
                \midrule
                Qwen3-8B \textit{full data}                                                     & 37.40 & 43.15 & 45.20 & 68.30 & 77.20 & 72.48 \\
                \midrule
                \ours-DataSelect \textit{random}                                    & 36.85 & 42.60 & 44.80 & 67.90 & 76.50 & 71.93 \\
                \ours-DataSelect \textit{STC}                          & 38.12 & 43.50 & 45.10 & 69.20 & 78.10 & 73.38 \\
                \ours-DataSelect \textit{STC + RCD}             & 39.29 & 44.23 & 45.32 & 70.57 & 79.23 & 74.64 \\
                \midrule
                \ours-8B \textit{STC + RCD + GCoT}                                & \textbf{40.25} & \textbf{45.30} & \textbf{46.12} & \textbf{71.40} & \textbf{81.15} & \textbf{75.95} \\
                \bottomrule
        \end{tabular}
    }
\caption{Ablation study of cold-start SFT with guided chain-of-thought. \textbf{STC} denotes stratified topic coverage, \textbf{RCD} denotes relevance category density, and \textbf{GCoT} denotes guided chain-of-thought rationale supervision. All \ours-DataSelect rows use one third of the training corpus, whereas the first row uses the full corpus and serves as the budget reference. The best value in each column is shown in bold.}
\label{tab:Ablation_cot}
\end{table}
\begin{table}[t]
  \centering
\resizebox{\linewidth}{!}{
      \begin{tabular}{l|ccc|ccc}
                \toprule
                \multirow{2}{*}{\textbf{Settings / Reward Mix}} &
                \multicolumn{3}{c|}{\textbf{Fine-grained Relevance Hierarchy}} &
                \multicolumn{3}{c}{\textbf{Coarse-grained Alignment}} \\
                & \textbf{Macro F1} & \textbf{Weighted F1} & \textbf{Accuracy}
                & \textbf{Precision} & \textbf{Recall} & \textbf{F1} \\
                \midrule
                \texttt{w/o} GRPO \textit{first-stage SFT}   & 40.25 & 45.30 & 46.12 & 71.40 & 81.15 & 75.95 \\
                \midrule
                GRPO \textit{Att + List}                     & 41.10 & 46.25 & 47.18 & 71.95 & 82.10 & 76.69 \\
                GRPO \textit{Att + List + Rule}              & 41.23 & 47.00 & 47.40 & 71.28 & 83.19 & 76.78 \\
                \midrule
                GRPO \textit{Att + List + Rule + Margin}     & \textbf{42.15} & \textbf{48.20} & \textbf{49.35} & \textbf{73.15} & \textbf{84.30} & \textbf{78.33} \\
                \bottomrule
        \end{tabular}
    }
\caption{Ablation study of components within the discriminative GRPO framework. \textbf{Att} denotes the attribute-wise reward, \textbf{List} denotes the list-wise reward, \textbf{Rule} denotes the rule-consistency reward, and \textbf{Margin} denotes the margin-based auxiliary loss on the terminal relevance token. All rows start from the same first-stage SFT checkpoint. The best value in each column is shown in bold.}    
\label{tab:Ablation_grpo}
\end{table}
\begin{table}[t]
  \centering
\resizebox{\linewidth}{!}{
      \begin{tabular}{l|ccc|ccc}
                \toprule
                \multirow{2}{*}{\textbf{Models}} &
                \multicolumn{3}{c|}{\textbf{Fine-grained Relevance Hierarchy}} &
                \multicolumn{3}{c}{\textbf{Coarse-grained Alignment}} \\
                & \textbf{Macro F1} & \textbf{Weighted F1} & \textbf{Accuracy}
                & \textbf{Precision} & \textbf{Recall} & \textbf{F1} \\
                \midrule
                \texttt{w/o} dual-head \textit{AR}   & 42.15 & 48.20 & 49.35 & 73.15 & 84.30 & 78.33 \\
                \midrule
                dual-head \textit{only-class-head}   & 41.25 & 47.80 & 48.95 & 71.12 & 83.95 & 77.00 \\
                dual-head \textit{one-dimensional}   & 42.98 & 49.32 & 50.65 & 72.59 & 85.73 & 78.61 \\
                dual-head \textit{multi-dimensional} & \textbf{43.68} & \textbf{50.13} & \textbf{50.98} & \textbf{74.82} & \textbf{86.45} & \textbf{80.22} \\
                \bottomrule
        \end{tabular}
    }
\caption{Ablation study of the dual-head distillation design. \textbf{AR} denotes autoregressive rationale decoding at inference time, \textit{only-class-head} removes the auxiliary generative head entirely, \textit{one-dimensional} keeps a single overall relevance head, and \textit{multi-dimensional} expands it into five sub-dimensional heads. All rows share the same backbone and the same distillation corpus. The best value in each column is shown in bold.}
\label{tab:Ablation_dual_head}
\end{table}
\begin{table}[!htbp]
  \centering
  \setlength{\tabcolsep}{3.5mm}{
      \begin{tabular}{l|cccc}
            \toprule
            \textbf{Models} & \textbf{HSR} & \textbf{LSR} & \textbf{CS} & \textbf{RS} \\
            \midrule
            Baseline   & --      & --         & --    & --      \\
            \ours-1.7B & +1.77\% & $-$3.22\%  & +2.49 & +3.63\% \\
            \bottomrule
        \end{tabular}
    }
\caption{Online A/B test improvements of the deployed \ours-1.7B student over the production baseline.}
\label{online_performance}
\end{table}

\subsection{Ablation Study}

\subsubsection{Cold-Start SFT with Guided Chain-of-Thought}

Table~\ref{tab:Ablation_cot} isolates the two ingredients of the first stage, namely sample selection and rationale supervision. Under a fixed budget of one third of the corpus, random sampling degrades every metric relative to full-data training, which confirms that the data scale itself is restrictive. Stratified topic coverage, denoted STC, improves Macro F1 from 36.85 to 38.12 at the same budget, surpassing the full-data baseline of 37.40, and balancing relevance category density, denoted RCD, adds a further increase to 39.29. The selected subset therefore exceeds full-data training by 1.89 points of Macro F1 and by 2.16 points of coarse F1 while consuming one third of the annotated data. These results demonstrate that combining curated high-quality data with explicit reasoning is significantly more effective than training on massive, unrefined datasets. 

Adding guided chain-of-thought (GCoT) on top of the selected subset produces the largest single increment of this stage. The GCoT row uses only rationales that pass the CoT checker, so the reported gain reflects screened rather than raw synthesized targets. Macro F1 rises from 39.29 to 40.25, fine-grained accuracy rises from 45.32\% to 46.12\%, and coarse Recall rises from 79.23 to 81.15. Since the underlying samples are identical across the last two rows, this increment is attributable to rationale supervision.

\subsubsection{Discriminative GRPO with Margin Alignment}

Table~\ref{tab:Ablation_grpo} progressively ablates the reward composition and the auxiliary objective. Combining the attribute-wise reward with the list-wise reward improves every metric over the first-stage SFT model, raising fine-grained accuracy from 46.12\% to 47.18\% and coarse F1 from 75.95 to 76.69. This indicates that decomposed supervision over relevance attributes and intra-group ordering provides usable optimization signal beyond a single terminal correctness reward.

Adding the rule-consistency reward improves fine-grained quality while redistributing the coarse-grained error. Macro F1 moves from 41.10 to 41.23, Weighted F1 from 46.25 to 47.00, and accuracy from 47.18\% to 47.40\%. At the binary boundary, Recall rises from 82.10 to 83.19 whereas Precision falls from 71.95 to 71.28, so coarse F1 changes only marginally from 76.69 to 76.78. This behaviour is consistent with what the reward actually constrains. The rule-consistency term rewards agreement between the generated rationale and the annotation guideline at the level of individual relevance dimensions, and it does not supervise the binary relevant against irrelevant decision. Enforcing dimension-level consistency makes the policy more willing to assign a non-zero grade to borderline documents that satisfy at least one dimension, which raises Recall at the cost of Precision and leaves the aggregate coarse F1 nearly unchanged. Since the terminal objective of our task is fine-grained ordinal judgement, we retain this reward and address the remaining boundary ambiguity with the auxiliary margin loss.

The margin loss delivers the largest improvement of the second stage and also resolves the Precision loss discussed above. Fine-grained accuracy rises from 47.40\% to 49.35\%, Weighted F1 rises from 47.00 to 48.20, coarse Precision recovers from 71.28 to 73.15, and coarse F1 rises from 76.78 to 78.33. Because this objective explicitly separates the logits of adjacent relevance grades at the terminal decision token, the resulting gain is concentrated where the reward signal alone is weakest, namely on near-tie cases between neighbouring grades.

\subsubsection{Dual-Head Distillation}

Table~\ref{tab:Ablation_dual_head} ablates the dual-head design under a fixed backbone. Removing the auxiliary generative objective entirely and training a pure classifier, the \textit{only-class-head} row, reaches 48.95\% fine-grained accuracy, which is 0.40 points below autoregressive inference at 49.35\%. A classification head on its own therefore does not preserve all of the signal that rationale supervision provides, and the benefit of our design cannot be attributed to replacing generation with classification. Reintroducing the auxiliary generative head recovers and exceeds this gap, lifting fine-grained accuracy to 50.65\% in the \textit{one-dimensional} setting, and expanding the classifier into five sub-dimensional heads attains the best result in the table at 50.98\% fine-grained accuracy and 80.22 coarse F1.

The gain originates from rationale-supervised representation learning combined with multi-dimensional supervision, and the generative head functions as a training-time regulariser on the shared backbone rather than as a component executed at inference. Moreover, sub-dimensional supervision is not merely an auxiliary detail, since moving from a single relevance target to five sub-dimensional targets adds 0.33 points of fine-grained accuracy and 1.61 points of coarse F1 without changing the training data.

Overall, we have verified the positive contributions of all components across the three stages, demonstrating the effectiveness of each component.
\clearpage
\begin{wraptable}{r}{0.48\textwidth}

\centering
\small
\resizebox{\linewidth}{!}{
      \begin{tabular}{l|c|cc|c}
            \toprule
            \multirow{2}{*}{\textbf{Serving Path}} & \textbf{Decoding} & \textbf{Latency} & \multirow{2}{*}{\textbf{Speedup}} & \textbf{Fine} \\
            & \textbf{Mode} & \textbf{(s/pair)} & & \textbf{Acc.} \\
            \midrule
            \ours-1.7B \textit{student}           & single pass & --            & --                    & 45.80 \\
            \midrule
            \texttt{w/o} dual-head \textit{AR}    & multi-step  & 5.68          & 1.00$\times$          & 49.35 \\
            \midrule
            dh \textit{only-class-head}    & single pass & \textbf{0.71} & \textbf{8.00}$\times$ & 48.95 \\
            dh \textit{one-dimensional}    & single pass & 0.82          & 6.93$\times$          & 50.65 \\
            dh \textit{multi-dimensional}  & single pass & 0.97          & 5.86$\times$          & \textbf{50.98} \\
            \bottomrule
        \end{tabular}
    }
\caption{Serving cost of the dual-head design. Latency is measured per query-document pair on 64 $\times$ NVIDIA H20 GPUs. The best value in each column is shown in bold.}
\label{tab:cost}
\end{wraptable}
\subsection{Cost Analysis}
\label{sec:cost}

Table~\ref{tab:cost} reports the serving cost of the same configurations, so that accuracy and cost can be read against each other. Autoregressive rationale generation costs 5.68 seconds per query-document pair, which is prohibitive for a production ranking stage that must score tens of candidates within a single request budget. Discarding the generative head at serving time reduces this to 0.97 seconds for the deployed multi-dimensional variant, which corresponds to a speedup of 5.86 times and a latency reduction of 82.9\%, while fine-grained accuracy improves from 49.35\% to 50.98\%.

Latency increases only modestly as classification heads are added, from 0.71 seconds to 0.82 seconds and then to 0.97 seconds. All heads share a single backbone forward pass, and each additional head introduces only a linear projection over the final hidden state. Compared with the \textit{only-class-head} variant, the multi-dimensional configuration gains 2.03 points in fine-grained accuracy at an additional cost of 0.26 seconds per pair. More importantly, the complete deployed model has 82.9\% lower latency than the autoregressive alternative. The additional classification heads therefore recover useful task-specific signals with limited marginal computation, resulting in a better accuracy-latency balance for the overall online ranking pipeline. This combination of higher predictive performance, lower serving latency, and reduced computational demand makes the multi-dimensional configuration a practical and valuable operating point for production deployment.

Retaining the generative head also preserves the language model's generation capability and provides an analytical interface that is unavailable in a pure classifier. The generative head is disabled during online ranking and therefore introduces no serving overhead, but it can be activated offline for case studies, error analysis, and diagnosis of individual predictions. Although online serving never invokes it, it provides useful qualitative evidence for understanding model behavior and tracing failure cases.

\subsection{Online Performance}
To evaluate the real-world impact of our framework, we deploy the distilled \ours-1.7B student and conduct a large-scale A/B test on the Yuanbao AI search application. 

High Satisfaction Rate (HSR) measures the prevalence of results that substantially satisfy the search intent, whereas Low Satisfaction Rate (LSR) captures clearly unsatisfactory results~\citep{jiang2015understanding}; higher HSR and lower LSR are preferred. Let $\mathcal{B}$ denote the set of evaluated search instances and let $y_i\in\{0,1,2,3,4,5\}$ be the satisfaction grade assigned to instance $i$. We define the HSR and LSR as:
\begin{equation}
\begin{aligned}
    \mathrm{HSR} &= \frac{1}{|\mathcal{B}|}\sum_{i\in\mathcal{B}}\mathbb{I}[y_i\in\{3,4,5\}],\\
    \mathrm{LSR} &= \frac{1}{|\mathcal{B}|}\sum_{i\in\mathcal{B}}\mathbb{I}[y_i\in\{0,1\}].
\end{aligned}
\end{equation}
Closed-set Score (CS) evaluates the model's ability to assess and rank retrieved web pages across multiple quality dimensions. The evaluation pipeline is a complete black box, without access to its internal retrieval or ranking mechanisms. This setting minimizes interference from model-specific preferences and internal signals, providing a more objective and comparable assessment of the model's evaluation. Relative Satisfaction (RS) measures how thoroughly the retrieved results satisfy the user's information needs. It captures the relationship between result relevance and the overall search experience. Because RS is primarily derived from the behavior and feedback of online users, it reflects real-world user preferences.

As shown in Table~\ref{online_performance}, \ours~improves every core business metric. The High Satisfaction Rate over grades 3 to 5, which reflects the proportion of queries whose intent is well satisfied by top-ranked results, increases by 1.77\%, and Relative Satisfaction, which quantifies the overall comparative improvement in user preference over the baseline system, increases by 3.63\%. The Low Satisfaction Rate over grades 0 to 1 decreases by 3.22\%, and the internal Closed-set Score improves by 2.49.

Two aspects of these results are worth noting. First, the direction of the offline and online conclusions agrees, since the model with higher fine-grained ordinal accuracy also produces fewer low-satisfaction impressions online, which supports our use of the fine-grained hierarchy as the primary offline objective. Second, the gains are obtained by a 1.7B non-autoregressive student rather than by the 8B teacher, which shows that the classification-only serving path is compatible with the latency constraints of live traffic while still transferring the benefit of reasoning supervision.

\section{Conclusion and Future Work}
In this paper, we introduced LARF, a three-stage framework bridging LLM reasoning with industrial search relevance estimation. By integrating cold-start SFT with guided CoT, discriminative GRPO with margin alignment, and dual-head distillation, LARF improves fine-grained ordinal relevance accuracy, sharpens ambiguous decision boundaries between adjacent grades, and removes autoregressive rationale decoding from the online serving path. It attains the strongest offline results among the compared baselines and improves user satisfaction in a large-scale commercial AI search system. Future work will explore multi-modal relevance judgments, shorter teacher rationales, and online reinforcement learning via real-time user feedback to build a self-improving ecosystem.

\bibliographystyle{ACM-Reference-Format}
\bibliography{custom}
\clearpage
\appendix

\section{Appendix}

\begin{minipage}{\textwidth}

\begin{PromptBox}{Prompt for Search Relevance Evaluation}
\footnotesize

\textbf{[System]}\\
You are a Search Result Satisfaction Evaluation Expert. Based on the
provided information, evaluate the search result across four dimensions:
\textbf{Semantic Relevance}, \textbf{Timeliness},
\textbf{Source Authority}, and \textbf{Content Quality}. You should also
predict the \textbf{Query Dimension} and assign an overall
\textbf{Satisfaction} score ranging from 0 to 5.

When scoring, comprehensively consider contextual information such as the
query type, timeliness sensitivity, and required authority level. The
relative importance of the four dimensions may vary depending on the query.
After assigning the scores, provide a detailed explanation supporting your
judgment.

\medskip
\textbf{[Input Information]}
\begin{itemize}
    \setlength{\itemsep}{0pt}
    \setlength{\parskip}{0pt}
    \setlength{\parsep}{0pt}
    \item \textbf{Query:} \texttt{\{query\}}
    \item \textbf{Document Title:} \texttt{\{title\}}
    \item \textbf{Document Content:} \texttt{\{content\}}
    \item \textbf{Search Time:} \texttt{\{search\_time\}}
    \item \textbf{Document Publication Time:} \texttt{\{publish\_time\}}
\end{itemize}

\textbf{[Auxiliary Information]}
\begin{itemize}
    \setlength{\itemsep}{0pt}
    \setlength{\parskip}{0pt}
    \setlength{\parsep}{0pt}
    \item \textbf{Query Domain/Topic:} \texttt{\{topic\}}
    \item \textbf{Publishing Platform:} \texttt{\{site\}}
    \item \textbf{Platform Name:} \texttt{\{sitename\}}
    \item \textbf{Platform Domain/Topic:} \texttt{\{site\_topic\}}
    \item \textbf{Platform Level:} \texttt{\{platform\_level\}}
    \item \textbf{ICP License Type:} \texttt{\{icp\_license\}}
    \item \textbf{Author Name:} \texttt{\{author\_name\}}
    \item \textbf{Author Domain/Topic:} \texttt{\{author\_topic\}}
    \item \textbf{Author Level:} \texttt{\{author\_level\}}
\end{itemize}

\textbf{[Constraint]}\\
Assess the document with respect to Semantic Relevance, Timeliness, Source
Authority, and Content Quality. Then determine the overall Satisfaction
score based on the above information. Output the dimension-wise assessments
and detailed explanation first, followed by the final Satisfaction score.

\end{PromptBox}

\captionof{figure}{
Prompt template for multi-dimensional search relevance evaluation.
}
\label{fig:relevance_prompt}

\vspace{1em}
\end{minipage}

\subsection{Prompt}
\label{app:prompt}

In this section, we introduce the prompts used for evaluating search
relevance. As shown in \figref{fig:relevance_prompt}, the model receives
metadata regarding the query-document pair and outputs scores based on
specific, predefined dimensions.

\subsection{Implementation Details}
We employ the Qwen3-235B-A22B-Instruct as the backbone for constructing \ours. All training experiments are conducted on 64 $\times$ NVIDIA H20 GPUs, with a batch size of 4 and a learning rate of $1\times10^{-5}$. The maximum training sequence length is set to 8192 tokens, and the model is trained for 2 epochs. For GRPO training, we use a rollout size of 8 and train for 2 epochs to mitigate the risk of overfitting. For inference, we employ vLLM with the specified serving configurations and set the temperature to 0 to ensure deterministic outputs.
\end{document}